\documentclass[oneside, a4paper, onecolumn, 11pt]{article}
\usepackage[left=2.5cm,top=2.5cm,bottom=2.5cm,right=2.5cm]{geometry}
\usepackage{amsmath}
\usepackage{graphicx}
\usepackage{fancyhdr}
\usepackage{calc}
\usepackage{amssymb}
\usepackage[super,sort]{natbib}
\usepackage{setspace}
\usepackage{amsfonts}
\usepackage{commath}
\usepackage[innercaption]{sidecap}
\usepackage[framemethod=TikZ]{mdframed}
\usepackage{hyperref}
\usepackage{parskip}
\usepackage{pifont}
\usepackage{sectsty}
\usepackage{times}
\sectionfont{\color{blue}}
\subsectionfont{\color{blue}}

\newcommand{\Rom}[1]{\expandafter\@slowromancap\romannumeral #1@}
\makeatother
{ \normalfont}

\newcommand{\m}[3]{#1_{#2 #3}}
\newcommand{\av}[1]{\langle #1 \rangle}

\expandafter\def\expandafter\normalsize\expandafter{%
	\normalsize
	\setlength\abovedisplayskip{0pt}
	\setlength\belowdisplayskip{5pt}
	\setlength\abovedisplayshortskip{0pt}
	\setlength\belowdisplayshortskip{5pt}
}

\definecolor{Gray}{gray}{0.75}

\newmdenv[backgroundcolor=Gray, leftmargin = 0pt, rightmargin = 0pt, linewidth = 0pt, roundcorner = 2 pt, innerleftmargin=5pt, innerrightmargin=5pt, innertopmargin=5pt, innerbottommargin=5pt]{Frame}

\begin{document}

	\linespread{1.2}

	\title{\huge  \color{blue}Mechanistic description of spontaneous loss of memory persistent activity based on neuronal synaptic strength}
	
	\author{Hillel Sanhedrai$^{1}$, Shlomo Havlin$^{1}$ \& Hila Dvir$^{1*}$}
	
	\maketitle

	\small{
		\begin{enumerate}
			\item
			\textit{Department of Physics, Bar-Ilan University, Ramat-Gan, Israel}			
		\end{enumerate}
		
		\begin{itemize}
			\item[\textbf{*}]
			\textbf{Correspondence}:\ \textit{dvirhila@gmail.com (HD)}
		\end{itemize} 
	}
	
	\vspace{4mm}
	
	\section*{Abstract}
	\textbf{
		Persistent neural activity associated with working memory (WM) lasts for a limited time duration. Current theories suggest that its termination is \textit{actively} obtained via inhibitory currents, and there is currently no theory regarding the possibility of a \textit{passive} memory-loss mechanism that terminates memory persistent activity. 
		Here, we develop an analytical-framework, based on synaptic strength, and show via simulations and fitting to wet-lab experiments, that passive memory-loss might be a result of an ionic-current long-term plateau, i.e., very slow reduction of memory followed by abrupt loss. We describe analytically the plateau, when the memory state is just below criticality. These results, including the plateau, are supported by experiments performed on rats. Moreover, we show that even just above criticality, forgetfulness can occur due to neuronal noise with ionic-current fluctuations, yielding a plateau, representing memory with very slow decay, and eventually a fast memory decay. 
		Our results could have implications for developing new medications, targeted against memory impairments, through modifying neuronal noise.
	}

\section{Introduction}

In many daily activities, such as maintaining telephone number digits, a small amount of information can be held and used even after the stimulus vanishes~\cite{zylberberg2017mechanisms}.
According to many physiological studies, the neural basis of this working memory (WM) ability, relates to persistent firing of the neurons~\cite{zylberberg2017mechanisms, wang2020macroscopic}.
The phenomenon, termed persistent activity, was recorded in animal models showing neurons with sustained elevated firing rates, after the stimulus disappeared~\cite{fuster1973unit, kubota1974visuokinetic, barak2010neuronal, zaksas2006directional}.
The mechanism of persistent activity is often associated with the intrinsic properties of the neurons themselves where ionic currents keep the spiking~\cite{egorov2002graded, fraser1996cholinergic}, or with connectivity within the neural circuit as a self-sustained network state~\cite{wang2020macroscopic, durstewitz2000neurocomputational}. 
However, it is not fully known, in particular for the self-sustained network state model, what finally terminates the persistent activity. While currently it is suggested that \textit{active} addition of external inhibitory currents terminate the persistent activity~\cite{durstewitz2000neurocomputational}, we suggest here a new mechanism where WM persistent activity is spontaneously terminated due to internal \textit{passive} mechanisms of either \textit{passive stochastic} or \textit{passive deterministic} environments.
Based on a self-sustained network state model, the Hopfield model~\cite{hopfield1982neural, hopfield1984neurons}, that describes the effect of synaptic strength on ionic-current, we provide an analytical description for spontaneous passive memory loss.

The self-sustained network state mechanism simulates WM persistent activity by analyzing the averaged firing rate, specifically analyzing the averaged firing rate vs. synaptic strengths~\cite{wang2020macroscopic, barbieri2008can, amit1997model}. 
Using those models, the persistent activity (WM) of a consistent firing rate could be achieved. However, describing the firing rate vs. synaptic strengths in a model is complex and cannot be studied analytically. In those models the average firing rate is obtained as the inverse of the average first passage time, and the equations are solved using numerical integration, hence analytical expressions cannot be formulated. In the present work, we base our analysis on an alternative representation of the WM model, that is on the \textit{ionic current} vs. synaptic strengths (instead of the \textit{firing rate}). 
Based on a simple current model, the Hopfield network model, we could perform a detailed mathematical analysis, which leads us to novel understandings of the persistent activity termination conditions.

The existing electrophysiological studies, using Hopfield network model, suggest that the mechanism to terminate WM is based on inhibition currents~\cite{amit1995learning, durstewitz2000neurocomputational}. In these studies the persistent activity is analyzed using a phase diagram of the \textit{current vs. the afferent stimulating current}. Using this modeling, the persistent activity appears as a current 
plateau, i.e. having approximately constant high values for a long time before it decays.
However, in order to terminate the plateau, that represent the loss of memory, an inhibitory current must be added to the model. Here we suggest natural mechanisms that terminate the persistent activity without using inhibitory currents.
In contrast to earlier studies, we propose here to analyze the behavior of the \textit{current vs. synaptic strengths} (instead of current vs. the afferent stimulating current). In this case our alternative approach leads us to develop a theory suggesting a new mechanism in which the WM persistent activity terminates \textit{passively}, and hence elucidates the yet unknown aspects of forgetfulness. Our results suggest that persistent activity shows a plateau behavior that ends spontaneously, without the need of active inhibition currents. Specifically, we suggest two spontaneous mechanisms for forgetfulness based either on: (i) deterministic dynamics properties when converging slowly from an unstable to a stable state. This we find to emerge in the Hopfield network model, causing the current to very slowly decrease (almost constant) in a plateau fashion, which finally decays to zero abruptly or (ii) natural neuronal noise that we introduce into the Hopfield network modeling, which we find to enable forgetfulness abruptly after a period of persistent activity.

Our WM analysis, is supported by results of electrophysiological experiments in young and old rats, measured by excitatory postsynaptic potentials (EPSP)~\cite{lynch2004long}, provoking long-term potentiation (LTP). LTP is a phenomenon that describes long lasting increase of the EPSP after stimulation. 
Many aspects of LTP are still unknown, and the common belief is that LTP is the physiological basis of memory that accompanies plasticity in synaptic connections~\cite{lynch2004long} - which are represented by the synaptic strength $\omega$ in our modeling.
There are many indications that LTP cannot act according to the same mechanism as long term memory (LTM) that lasts years, mainly as LTP never lasts more than couple of weeks (usually few days)~\cite{lynch2004long}. Thus, LTP is an excellent candidate having much in common with persistent activity, due to its persistent electrophysiological activity characteristics~\cite{martin2000synaptic}.
Moreover, we show resemblance between the dynamics of LTP and the persistent activity of Hopfield network model in Fig.~\ref{fig:plateau}~d and Fig.~\ref{fig:LT}~d.
While it is yet unknown how to describe the relation between the time duration of LTP experiments and average synaptic strength $\omega$,
our framework here suggests such analytical relationship and demonstrates it for aged and young rats as observed in the EPSP experiments.
Specifically, our calculation of $\omega$ suggests that LTP protocol, in both young and aged rats, produces weak synaptic strength.

\section{Theoretical background}

A common model of WM persistent activity is based on sustained excitation activity of highly connected neuronal cells. 
This concept was introduced in the Hopfield model~\cite{hopfield1982neural, hopfield1984neurons}, which describes the variables as the neuronal membrane potential~\cite{hopfield1984neurons} or current~\cite{amit1995learning, durstewitz2000neurocomputational}. The WM persistent activity model was formalized as~\cite{amit1995learning, durstewitz2000neurocomputational},

\begin{equation}
\tau \dod{I_i}{t} = -I_i + \sum_{j \neq i} \m wij \ln \left( \frac{I_j}{C} \right) \theta\left(\frac{I_j}{C}-1 \right) + I_{\rm Aff},
\label{eq:model}
\end{equation}
where $I_i$ represents the current in neuron $i$ and $\tau$ is the time constant. 
Here, $\theta(x)$ is the Heaviside step function, determining an excitation threshold $C$, where only above $C$ the current of neuron $j$ impacts neuron $i$. $I_{\rm Aff}$ is the external current excitation, which is obtained by an external stimulation (e.g., sensory stimulation). 
In Eq.~\eqref{eq:model}, $w_{ij}$ represents the interaction strength (synaptic strength) between neurons $i$ and $j$, constructed as a complete graph (each neuron is connected to all neurons). 

\begin{figure}[h!]
	\centering
	\includegraphics[width=0.5\linewidth]{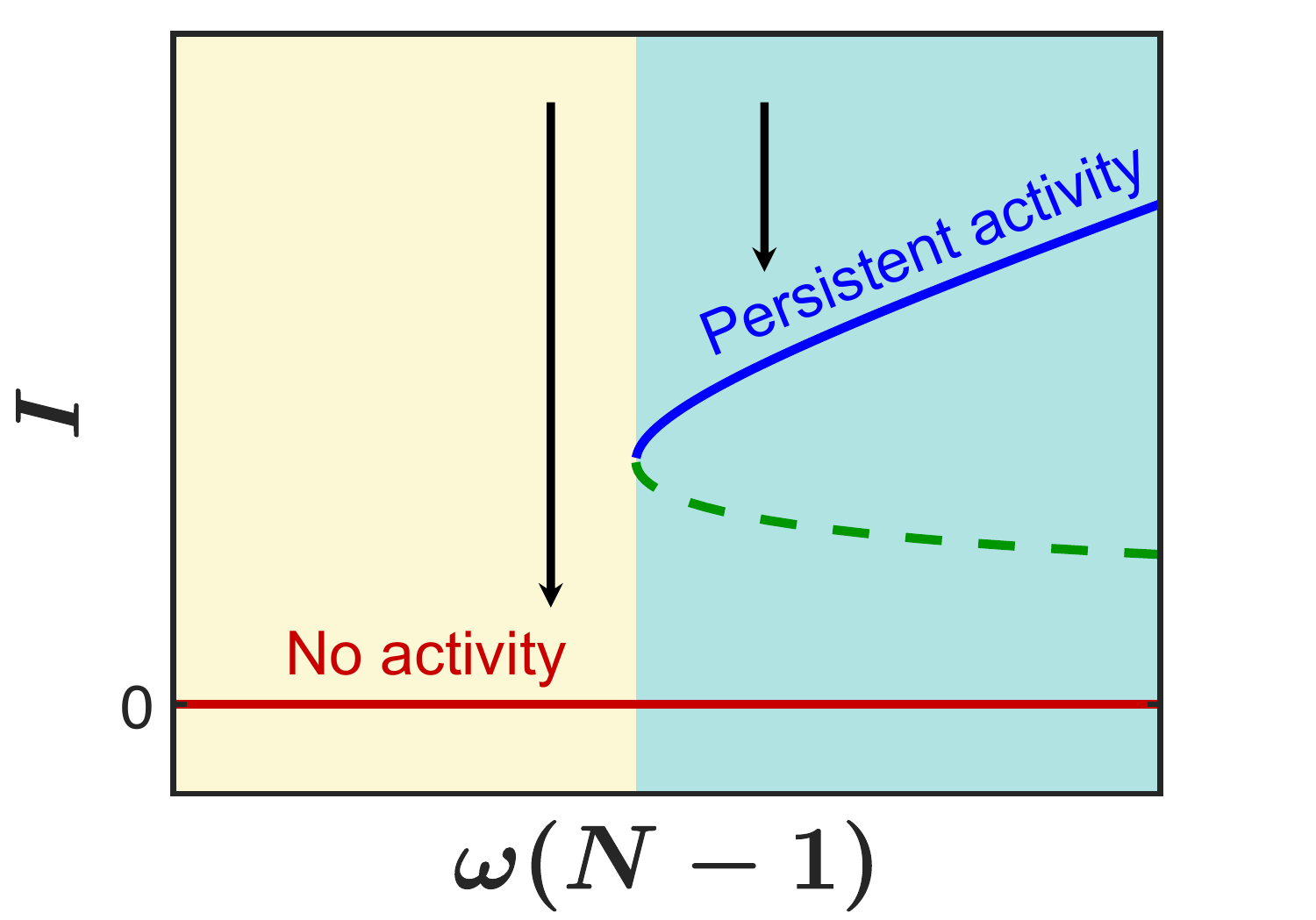}	
	\caption{\footnotesize{\bf Memory phase diagram.}
		The phase diagram in $\omega$-space obtained from Eq.~\eqref{eq:MF}. The blue and the red continuous lines represent the steady states solutions, while the dashed green line represents an unstable solution separating between the two basins of attraction.  
		Note that according to Eq.~\eqref{eq:omega}, the value of $\omega(N-1)$ is actually the mean synaptic weight.
		See also the fixed points and stability analysis in Fig.~S3 (Supplementary).
	}
	\label{fig:phaseDiagram}
\end{figure}

We hypothesize here that the weights $w_{ij}$ vary with a value that is imprinted by the neurons' individual characteristics. Moreover, we assume that $w_{ij}$ depends also upon the strength of the external stimulation, specifically, high external input yields high neuronal connections $w_{ij}$.
Therefore, our analysis here will be, in contrast to earlier approaches, based on a memory phase diagram in the $\omega$-space as shown in Fig.~\ref{fig:phaseDiagram}, where $\omega$ is the average over all weights $w_{ij}$.

Next, we analyze Eq.~\eqref{eq:model} using a Mean-Field (MF) approximation~\cite{gao2016universal} by replacing the varying interaction term by an averaged quantity, and by that we obtain a single equation for the average current (see Supplementary Section S.2). 
Specifically, in order to execute complex tasks such as memory, the brain neurons become highly connected~\cite{solomon2003complexity, ben2004complex, gosak2018network}, with networking properties close to a complete graph, which makes the MF solution more accurate. 
The obtained single MF equation is

\begin{equation}
\tau \dod{I}{t} = -I + \omega (N-1) \ln \left( \frac{I}{C} \right) \theta\left(\frac{I}{C}-1 \right) + I_{\rm Aff},
\label{eq:MF2}
\end{equation}
where 
\begin{equation}
I = \dfrac{1}{N}  \sum_{j=1}^{N} I_{j},
\label{eq:meanI}
\end{equation}
and	with defining $\omega$ as
\begin{equation}
\omega = \dfrac{1}{N(N-1)}  \sum_{j \neq i} \m wij.
\label{eq:omega}
\end{equation}

Thus, $I$ is the average current across the neurons (Eq.~\eqref{eq:meanI}), and $\omega$ is the average interaction weight between neurons (Eq.~\eqref{eq:omega}).

Concerning memory, it is of importance to study the sustained current when $I_{\rm Aff}=0$, using,

\begin{equation}
\tau \dod{I}{t} = -I + \omega (N-1) \ln \left( \frac{I}{C} \right)\theta\left(\frac{I}{C}-1 \right) .
\label{eq:MF}
\end{equation}

We next track the fixed points of Eq.~\eqref{eq:MF} by setting $\dif I/ \dif t = 0$, which yields three solutions for the average current $I$ as a function of the average weight $\omega$: two stable solutions and one unstable solution.
The $\omega$-space splits into two areas. 
Indeed, Fig.~\ref{fig:phaseDiagram} (see also Fig.~S3 in Supplementary) shows exactly this: small values of $\omega$ yield a single dormant state (red), whereas large $\omega$ yields three fixed points (including two steady states, active (blue) and dormant (red)). We assume that the active state enables the existence of a persistent activity (i.e., WM) for large enough $\omega$, while for small $\omega$ the memory cannot stay for long (no-activity).

Next we find the critical (tipping) point ($\omega_c,I_c$) of the transition from no-activity to persistent activity, while noticing that this point is a junction of a stable fixed point (continuous blue line in Fig.\ \ref{fig:phaseDiagram}) and an unstable one (dashed green line). 
This requirement yields (see Supplementary Section S.4),
\begin{equation}
\omega_c = \dfrac{e\cdot C}{N-1}, \quad
I_c=e\cdot C.
\label{eq:wcIc}
\end{equation}

We conclude, from Eqs.~\eqref{eq:MF2}~-~\eqref{eq:wcIc}, that $\omega>\omega_c$ is the condition for the creation of persistent activity, otherwise the memory is not preserved and the current decays to zero, hence the range of $\omega<\omega_c$ is a non-activity region.

In the following analyses we explore the impact of the interaction (synaptic) weights on the fundamental temporal memory characters through theoretical and simulation results as well as experimental wet-lab studies.
The experimental wet-lab relates our theory to hippocampal of anesthetized young and aged rats model~\cite{lynch2004long}. Note that using hippocampal model is a widely accepted paradigm of simple brainlike structures~\cite{stoop2021excess}.

\section{Results}

Next we specify and show results that support our hypothesis about the possible mechanisms of \textit{passive} persistent activity forgetfulness. Our suggestion, based on solutions of Eq.~\eqref{eq:MF} (Fig.~\ref{fig:phaseDiagram}), for two mechanisms are: (i) A current flow that shows a natural long-term plateau with very slow decay under certain conditions. The memory is represented as an almost constant current for a limited time followed by a natural fast decay to zero, (ii) Neuronal noise fluctuations of the ionic-currents that lead to its fast decay, with persistent activity duration that is described by a  \textquoteleft first passage time\textquoteright{} analysis.

\subsection{Passive deterministic forgetfulness: Deterministic plateau}

Here we describe the passive mechanism for loss of persistent activity based on deterministic dynamic analysis, that we find to emerge in the transition from unstable to stable state in the Hopfield model, Eq.\ \eqref{eq:MF2}. 
Before explicitly presenting our theory regarding {memory persistent activity} forgetfulness, note that an intuitive explanation and demonstration is provided in Supplementary section Fig.~S1.

\begin{figure}[h!]
	\centering
	\includegraphics[width=0.99\linewidth]{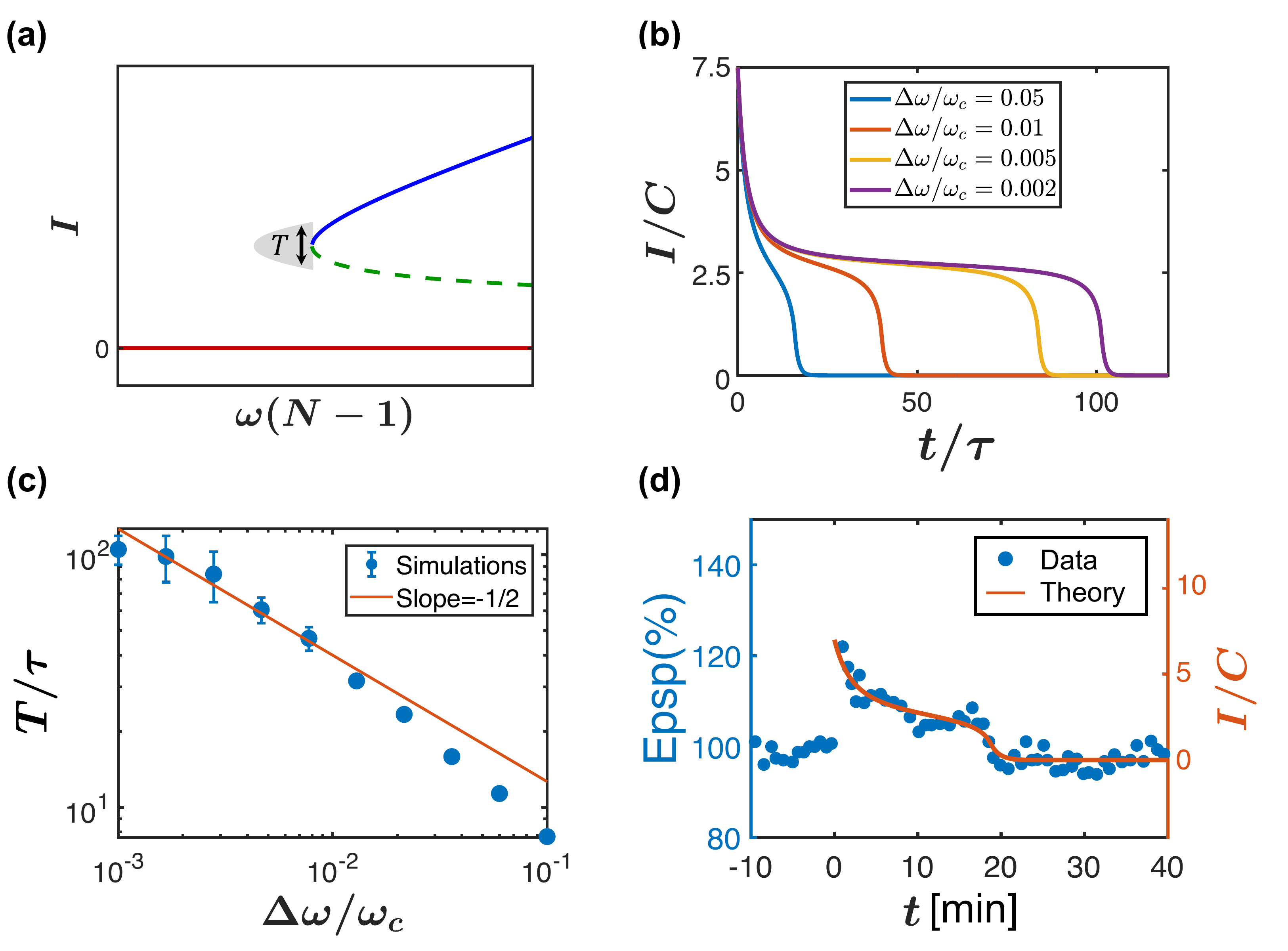}
	
	\caption{\footnotesize {\bf Deterministic plateau.} 
		(a) Deterministic scenario describing current just below the criticality. Due to the closeness to the fixed point, the current decreases towards zero very slowly, such that a pseudo memory emerges. 
		(b) Below criticality, $\omega<\omega_c$, the average current decays to zero, however close to criticality there is a long slow decay like a plateau, which is similar to a steady state. This quasi-steady state  plateau increases its time length, plateau duration, when simulated for smaller values of $\Delta\omega/\omega_c$.
		(c) The duration of the plateau, $T$, scales as $\Delta\omega^{-1/2}$ such as predicted by Eq.\ \eqref{eq:Tplateau}. 
		Note that the decay of $T$ with an exponent $-1/2$ is analogous to critical interdependent networks~\cite{zhou2014simultaneous, gao2022introduction}.
		Here we show averaged values of simulations over $10$ realizations assuming Gaussian distribution for the synaptic weights. 
		Simulations in panels (b) and (c) were performed with $\tau=1$, $C=2$ and $N=100$ neurons, (i.e., without Mean-Field (MF) approximation).
		(d) Experimental data of anesthetized aged rats~\cite{lynch2004long} (blue dots), shows a very good agreement with our theoretical simulation results (red line) for our suggested state variable $\Delta\omega/\omega_c=0.04$ (i.e., $\omega=0.96\omega_c$). 
		All stimulations use constant values of: $I_0=14$, $C=2$ and  $\tau=1$~[min] (as assumed in sleep brain models~\cite{dvir2018neuronal}). The simulation results are shown for normalized $I/C$ and thus have no units. This is since different scales of $C$, and hence of $I$, have no affect on the temporal dynamics but only their ratio.
		The EPSP units, in panel (d) and in the next Fig.~\ref{fig:LT}~d, are shown in percentage~\cite{lynch2004long}, which indicate increase in EPSP slope relative to pre-stimulation baseline.
		}
	\label{fig:plateau}
\end{figure}

We describe the passive mechanism for forgetfulness, based on a plateau formation shape of the current, which takes place near criticality just below the critical point.
Specifically, we simulate the current in the region (gray area in Fig.~\ref{fig:plateau}~a) of a weak synaptic strength, which includes only the dormant steady state. However, due to its closeness to the persistent-activity area, the forgetting process is slow, and might include a long plateau which acts as a quasi-steady state, for some period.
This means that, in the $\omega$-space, this plateau scenario is for $\omega<\omega_c$, i.e., $\omega=\omega_c-\Delta\omega$, therefore $I$ converges to zero, but since $\omega$ is near criticality ($\Delta\omega$ is small) the current decays very slowly in a plateau shape, 
see Fig.\ \ref{fig:plateau}~a~-~b (see also the analogous plateau phenomena in percolation of interdependent networks~\cite{zhou2014simultaneous, gao2022introduction}).
The duration of this slow decaying plateau depends on the proximity of $\omega$ to $\omega_c$, i.e., on $\Delta\omega$, as depicted next (based on \cite{strogatz2018nonlinear}). 
In order to analytically estimate the time duration of the plateau, $T$, we focus on the region around the plateau formation, that is: $\omega \to \omega_c$ and current close to $I_c$, i.e., the time duration where $\delta I=I-I_c$ is small.
In general, for two {such currents}, close to $I_c$, 
with $2\epsilon$ currents difference {between them}, their time interval, $T$, can be described as:
\begin{eqnarray}
	T = \int \dif t \approx \int_{I_c+\epsilon}^{I_c-\epsilon } \dod{t}{I} \dif I = \int_{I_c+\epsilon}^{I_c-\epsilon } \left(\dod{I}{t}\right)^{-1} \dif I = \tau\int_{-\epsilon}^{\epsilon} \left(-\tau\dod{\delta I}{t}\right)^{-1} \dif \delta I .
	\label{eq:T_general}
\end{eqnarray}
To calculate the argument of the integral, we analyze the behavior of Eq.\ \eqref{eq:MF} in the limit of $\omega \to \omega_c^-=e\cdot C(N-1)$ and $I\to I_c = e \cdot C$ (see Eq.\ \eqref{eq:wcIc}). 
We further use Eq.~S.26 (Supplementary) for small $\Delta\omega$, keeping also the second order term of $\delta I$,
\begin{equation}
	-\tau \dod{\delta I}{t} \approx \frac{I_c}{2} \left[ \left(\frac{\delta I}{I_c}\right)^2 + 2 \frac{\Delta\omega}{\omega_c} \frac{\delta I}{I_c} + 2 \frac{\Delta \omega}{\omega_c} \right] .
	\label{eq:dI_second_order}
\end{equation}
Placing Eq.~\eqref{eq:dI_second_order} into Eq.~\eqref{eq:T_general}, yields,
\begin{equation}
	T  \approx 2\tau \int_{-\infty}^{\infty} \frac{1}{x^2 + 2 b x + 2 b} \dif x = \frac{2\tau\pi}{\sqrt{b}\sqrt{2-b}} ,
	\label{eq:T0}
\end{equation}
where $x=\delta I/I_c$ and $b=\Delta\omega/\omega_c$. 
Note that in Eq.~\eqref{eq:T0} we replaced the integration limits to $\pm \infty$, since the integral is governed by small values of $x$ and large values do not change the leading term.

Since $b=\Delta\omega/\omega_c$ is small, we finally get the plateau duration,
\begin{equation}
	T  \approx \tau \frac{\sqrt{2}\pi}{\sqrt{\Delta\omega/\omega_c}} .
	\label{eq:Tplateau}
\end{equation}
We note that, according to our analysis, the deterministic plateau duration also depends on $\tau$. Hence environmental physiological conditions, such as sleep/anesthesia that affect $\tau$, will accordingly change the time scale of $T$. Moreover, note that Eq.~\eqref{eq:Tplateau} depends on the same parameters (though with different scalings) as Eq.~S.28 (Supplementary), which refers to the typical time decay of the current at conditions where there is no plateau formations.

Figure~\ref{fig:plateau}~c illustrates the good fit between the theory, Eq.~\eqref{eq:Tplateau} (red line), and the simulation results (blue dots). As expected, this fit is better at low values of $\Delta\omega$, i.e., when $\omega$ is closer to criticality.

Furthermore, a closer look on recent empirical studies, Fig.~\ref{fig:plateau}~d, performed with EPSP measuring memory of rats models~\cite{lynch2004long,grosser2020loss}, shows memory with a quasi-steady state long plateau similar to our theory and simulation results. Fig.~\ref{fig:plateau}~d compares between memory of aged rats~\cite{lynch2004long} (blue dots) and simulating the current with $\omega$ close to criticality ($\Delta\omega/\omega_c=0.04$). The excellent agreement between the plateau simulations and the EPSP measurements supports our theory and further suggests that the average synaptic strength with respect to the critical synaptic strength is for this case, $\omega=0.96\omega_c$. We note that similar plateau behavior has been also observed for EPSP measurements of memory in pilocarpine-treated rats~\cite{grosser2020loss}. 

\subsection{Passive stochastic forgetfulness: Stochastic plateau}

In the previous Section we obtained a plateau from a state just below criticality. In the present Section we show that a plateau can be formed also at a state just above criticality.
We will demonstrate, next, that for this scenario to take place, stochastic conditions must be present, as indeed occurs physiologically due to neuronal noise.
For that matter, we first analyze persistent activity without considering forgetfulness and analyze EPSP of young rats.
Specifically, we will show that by adding neuronal noise a plateau is formed even at certain conditions above criticality.

\subsubsection{Persistent activity analysis}

In this Section we study the regime in $\omega$-space (Fig.~\ref{fig:phaseDiagram}) of $\omega>\omega_c$, where persistent activity is maintained, and analyze the memory dynamic when converging to a steady state. 
We study here persistent activity, since we wish to better understand the effect of $\omega$ in young rats EPSP data, which is concluded at the end of the present section.
Specifically, we show that the current converges to its stable state exponentially with a typical time constant (representing the memory length) that depends on the average weight $\omega$.

For a steady state memory solution, the derivative of the current vanishes in Eq.~\eqref{eq:MF} and thus obtaining:
\begin{equation}
0 = -I_{\rm LT} + \omega(N-1) \ln (I_{\rm LT}/C) ,
\label{eq:fpLT}
\end{equation}
where $I_{LT}$ is the persistent current that yields the memory persistent activity (steady-state solution). Without intervention, such as the active inhibitor currents, this could have yielded a long term (LT) memory (LTM) solution.
Here we study this region of persistent activity, whereas in the next Section we study how it could be still terminated to forgetfulness.

Next we are interested in describing the dynamics of LTM using Eq.\ \eqref{eq:MF} in two different limits: (i) $I \gg I_{\rm LT}$ and (ii) when $I$ approaches $I_{\rm LT}$ (when $I \to I_{\rm LT}$).

(i) In the first limit ($I \gg I_{\rm LT}$), the behavior is trivial: since $I$ is very large it can be assumed that $I \gg \ln I$, so that Eq.~\eqref{eq:MF} can be written approximately as $\tau {\dif I}/{\dif t} \sim -I$, and therefore an exponential decay with a typical time $\tau$ is obtained, 
\begin{equation}
I(t) \sim \exp(-t/\tau).
\label{eq:tauBegining}
\end{equation}

\begin{figure}[h!]
	\centering
	\includegraphics[width=1.0\linewidth]{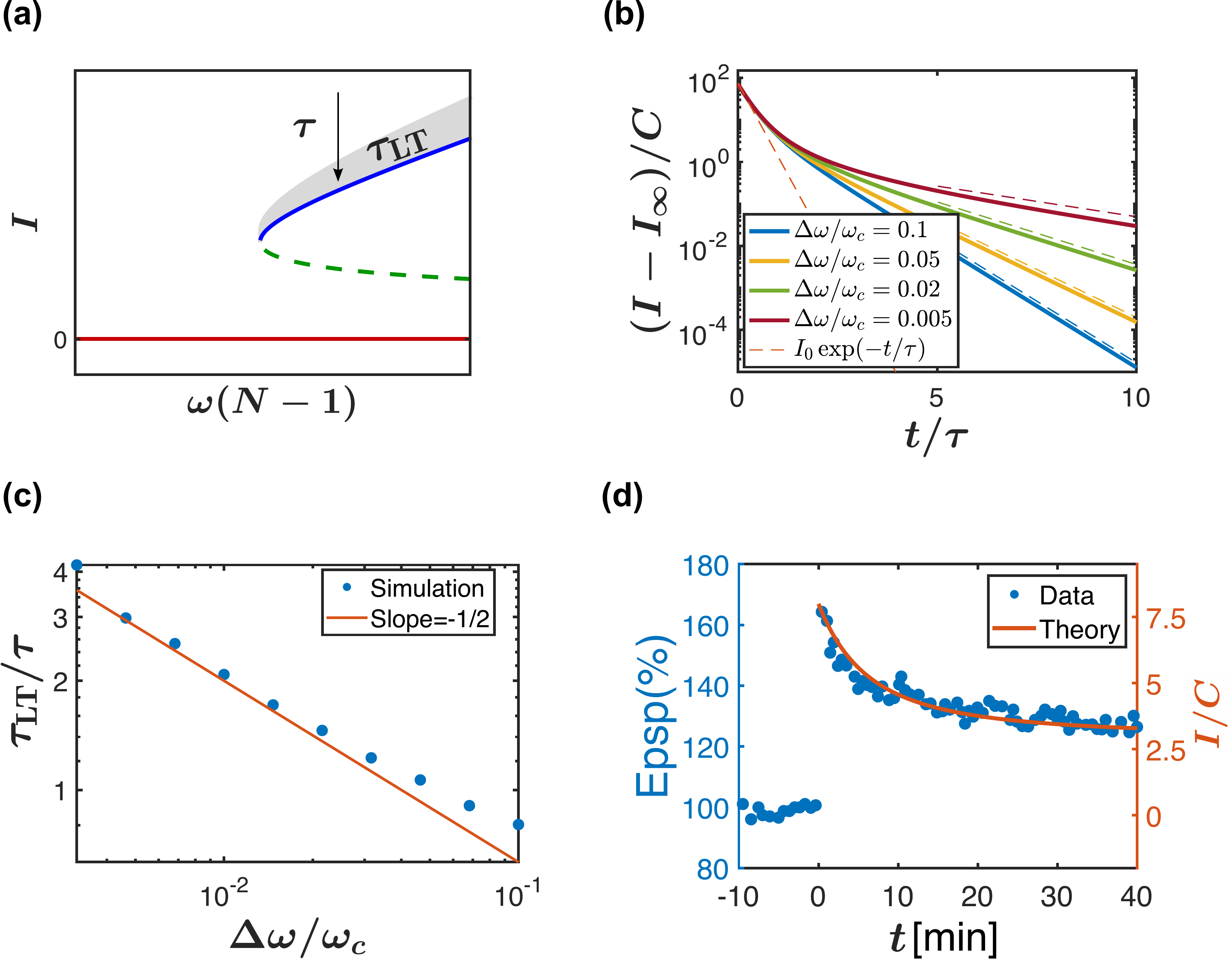}		
	\caption{\footnotesize{\bf Persistent activity temporal decay.} 
		(a) The current converges to a positive steady state because the average synaptic weight $\omega$ is above criticality $\omega>\omega_c$ and the system is in the active region. However, we identify two different typical time constants, long and short, for the decay of the memory. Far from the steady state there is a universal short time decay $\tau$ while close to the final state (gray area) there is a larger time decay $\tau_{\rm LT}$.
		(b) The decay of the average current $I$ with time is exponential with fast (at the beginning) and slow (towards the end) exponential decay. The fast and slow exponents represent characteristic time constants $\tau$ and $\tau_{\rm LT}$, respectively, and fit well to their theoretical values in Eq.~\eqref{eq:tauBegining} and Eq.~\eqref{eq:tauLT}, which are marked in dashed lines. 
		(c) Here we provide support for the analytical approximation, Eq.\ \eqref{eq:tauLTclose}, for small $\Delta\omega$, showing a power-law with exponent $-1/2$. 
		Simulations in panels (b) and (c) were obtained for a network of size $N=100$ neurons (i.e., without using Mean-Field (MF) approximation), threshold of $C=2$, $\tau=0.25$ and initial condition (by stimulating current $I_{\rm Aff}$) of $I_0=150$. 
		(d) Empirical memory data as collected from anesthetized young rats model~\cite{lynch2004long} (blue) compared to our model (red) with $C=2$, $\tau=3$, $I_0=16$ and $\Delta\omega/\omega_c=0.006$ (i.e., $\omega=1.006\cdot\omega_c$) yielding $\tau_{\rm LT}\approx27$~min. 	
		Note that the value of $\omega/\omega_c=1.006$ is normalized, and therefore is not affected by units or scaling parameters which affect both $\tau$ and $\tau_{\rm LT}$ (see  Eq.~\eqref{eq:tauLTclose}). Hence, while other parameters sets might yield same current temporal configuration, the value of $\omega/\omega_c$ is resilience and fixed per configuration.
	}
	\label{fig:LT}
\end{figure}

(ii) In the second limit, close to $I_{\rm LT}$ (when $I \to I_{\rm LT}$), we expand Eq.\ \eqref{eq:MF} as $I=I_{\rm LT}+\delta I$ where $\delta I$ is small (see details in Supplementary Section S.5), and obtain,
\begin{equation*}
\tau \dod{\delta I}{t} \approx -\bigg( 1 -\frac{1}{\ln (I_{\rm LT}/C)} \bigg) \delta I .
\end{equation*}
This implies an exponential decay with a typical time decay,
\begin{equation}
\tau_{\rm LT} = \frac{\tau}{1 -\frac{1}{\ln (I_{\rm LT}/C)}} .
\label{eq:tauLT}
\end{equation}
Note that this exponential decay is valid under the condition $\delta I / I_{\rm LT} \ll \ln (I_{\rm LT}/I_c)$ as aforementioned, namely within a region above the steady memory, see Fig.\ \ref{fig:LT}~a. The range of this validity region is getting narrower close to criticality.

Figure~\ref{fig:LT}~b shows results from simulating Eq.~\eqref{eq:model} with $I_{\rm Aff}=0$ for the LTM case, that is for $\omega=\omega_c+\Delta\omega$ and $\Delta>0$, showing the average current, $I$, temporal decay. 
Each colored line is for a different value of $\Delta\omega/\omega_c$. Particularly, for all values of $\Delta\omega/\omega_c$ two exponential decays can be observed. The first exponent, at short time scales, t,  (in the beginning of the simulations), is $\tau$, which appears for all $\Delta\omega/\omega_c$ values, in accordance to Eq.~\eqref{eq:tauBegining} (Fig.~\ref{fig:LT}~b dashed orange line). The second exponent, shows up at long time scales (towards the end of the simulation time and close to the steady state), representing a more moderate decay with an exponential value that fits well to the $\tau_{\rm LT}$  in Eq.~\eqref{eq:tauLT} (Fig.~\ref{fig:LT}~b, four dashed lines). Note that the first exponent, $\tau$, is smaller than the second exponent, $\tau_{\rm LT}$, which also fits well Eq.~\eqref{eq:tauLT} implying that indeed, $\tau_{\rm LT}>\tau$.

Next, we evaluate $\tau_{\rm LT}$ when $\omega$ is close to the critical value $\omega_c$ from above. To track the behavior in proximity of $\omega_c$ (for small $\Delta\omega$), we expand Eq.\ \eqref{eq:fpLT} using $\omega=\omega_c+ \Delta\omega$ and $I_{\rm LT} = I_c + \delta I_{\rm LT}$ (see Supplementary). Recalling that $I_c=e\cdot C$, and $\omega_c=e\cdot C/(N-1)$ (Eq.~\eqref{eq:wcIc}) we get
\begin{equation*}
\delta I_{\rm LT} \approx I_c \sqrt{2 \Delta\omega/\omega_c}.
\end{equation*}
Substituting this into Eq.\ \eqref{eq:tauLT}, we finally obtain,
\begin{equation}
\tau_{\rm LT}  \approx \frac{\tau}{\sqrt{2 \Delta\omega/\omega_c}}.
\label{eq:tauLTclose}
\end{equation}

This relation, Eq.~\eqref{eq:tauLTclose}, is in excellent agreement with our simulation results as illustrated in Fig.~\ref{fig:LT}~c. Fig.~\ref{fig:LT}~c summarizes the values of the time decay exponent, $\tau_{\rm LT}$, as obtained from the simulations of the current $I$ close to its steady state (the gray area in Fig.~\ref{fig:LT}~a and the straight lines close to steady-state in Fig.~\ref{fig:LT}~b). It is clearly seen that the relation between $\Delta\omega/\omega_c$ and $\tau_{\rm LT}$ is a power law with an exponent $-1/2$, supporting Eq.~\eqref{eq:tauLTclose}. Note, as expected, that for small values of $\Delta\omega/\omega_c$, i.e.\ when $\omega$ is close to $\omega_c$, the simulation results in Fig.~\ref{fig:LT}~c, are more accurately described by a $-1/2$ slope.

After investigating the model theoretically and verifying it by simulations, we explore the theoretical parameter $\omega$ and its relation to actual electrophysiological experiments of measuring memory data of EPSP. 
We used empirical EPSP data of long-term potentiation (LTP)~\cite{lynch2004long}, which is a phenomenon observed in brain areas involved in memory storage such as the hippocampus, and it describes long lasting increase of the EPSP after stimulation. The common belief is that LTP is the physiological basis of memory which accompanies plasticity in synaptic connections~\cite{lynch2004long}. 
Fig.~\ref{fig:LT}~d shows the LTP dynamics as experimentally measures in young rats~\cite{lynch2004long} (blue dots) compared to our theoretical model of current dynamics as simulated for $(\omega-\omega_c)/\omega_c=\Delta\omega/\omega_c=0.006$ (red line). 
The excellent agreement between the data and the model suggests indeed that the average weight $\omega$ relation to the critical weight $\omega_c$ is $\omega=1.006\cdot\omega_c$. Thus, in this data of anesthetized young rats, the average synaptic weight is fairly close to the critical LTM weight, and just above it. 
However, in this LTP memory data, after one week~\cite{lynch2004long} the memory eventually decays to zero, much faster than the common LTM. In the next Section we explain how this system could move from just above critical point to forgetfulness.

\subsubsection{Passive stochastic forgetfulness: Stochastic plateau - effect of neuronal noise}
To consider neuronal noise inflations, we extend the model in Eq.~\eqref{eq:MF2} so that 
it includes neuronal noise of normal distribution with std $\sigma$ and zero mean, using the stochastic differential equation:
\begin{eqnarray}
\tau \dod{I}{t} = -I + \omega (N-1)  \ln \left( \frac{I}{C} \right)\theta\left(\frac{I}{C}-1 \right) + \tau\cdot\sigma\cdot n(t),
\label{eq:OU_process}
\end{eqnarray}
where $n(t)$ is a white noise.
We assume that the stimulating current $I_0$, which initiates the memory, has terminated so that now the neurons are influenced by the neuronal noise that 
fluctuates around zero. 
Note that here we analyze the average current, which is averaged over all neurons, same as in the definition, Eq.~\eqref{eq:MF2}. 

Next, we ask: when will the average current $I$, which here evolves randomly, cross down a critical value (the threshold, $C$, in the average case), and consequently the system will collapse to the dormant state regime, i.e., fully forgetfulness state.
In fact, this phenomenon actually describes the problem known as the \textquoteleft first passage time\textquoteright{} of a stochastic process~\cite{redner2001guide}. 
Specifically, we wish to examine the time when the mean current $I$ crosses down for the first time the threshold $C$, and consequently decays to the forgetfulness state.
In order for this memory-loss scenario to occur, $\omega$ must have a value slightly above the critical $\omega_c$, so that the noise could push it towards the dormant state regime. This occurrence starts as persistent activity state, and terminated passively in the presence of neuronal noise.

\begin{figure}[h!]
	\centering
\includegraphics[width=0.99\linewidth]{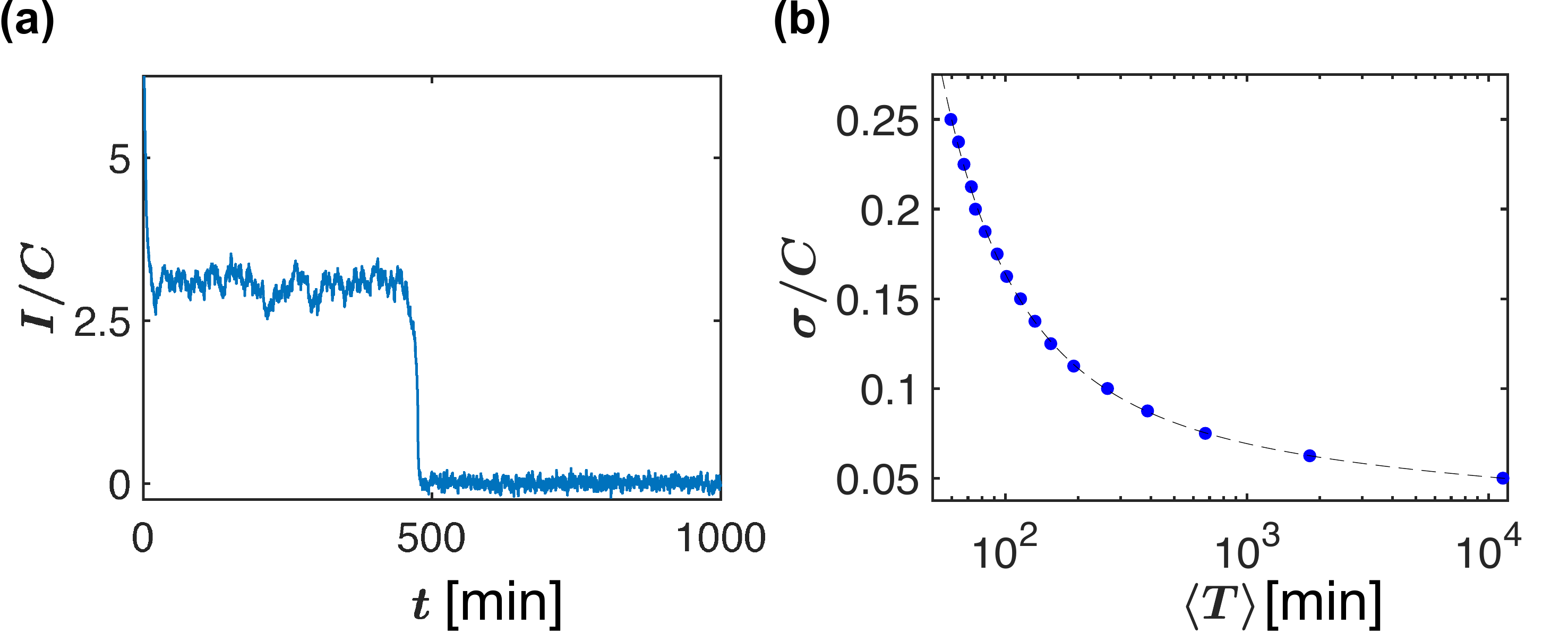}
	\caption{\footnotesize{\bf Stochastic plateau: Simulations of neuronal noise (white noise).}
		(a) In the persistent activity phase near criticality, there is a chance that afferent input noise will lead to crossing the criticality towards the forgetting region.
		The average current evolves in time according to Eq.~\eqref{eq:OU_process}. 
		Simulating the case $\Delta \omega/\omega_c=0.006$ and $\sigma=0.17$, 
		yields a first passage time $T$, where the current $I$ sharply decays below the threshold $C$, at around $T\approx 400$~[min]. 
		(b) We repeat the simulation of panel (a) for different values of noise level $\sigma$. For each $\sigma$ value, simulation has been repeated over $1,000$ realizations in order to obtain a mean value of $\av{T}$. 
		The average time until forgetting $\av{T}$ is marked in blue circles. As $\sigma$ decreases, the mean time duration $\av{T}$ increases inversely with $\av{T}$, see the dashed line. From this graph, given $\av{T}$ the noise level $\sigma$ can be calculated (the threshold $C$ value is generally known). 		
		All simulations have been performed with $\tau=1$~[min] (in accordance with sleep brain models~\cite{dvir2018neuronal}), $C=80 pA$~\cite{tateno2004threshold} and $N=100$. The units of $\av{T}$ are in [min], same as the units of $\tau$.
	}
	\label{fig:noise_direct}
\end{figure}

In Fig.~\ref{fig:noise_direct}~a we simulate Eq.~\eqref{eq:OU_process}, i.e., the stochastic plateau for values, $\sigma=0.17$ and $\omega=1.006\cdot \omega_c$, that is $\omega$ is slightly above $\omega_c$. 
Starting at persistent activity state close to the critical $\omega_c$, due to the 
neuronal noise fluctuations, the current $I$ rapidly drops to zero at around $T\approx 400$~[min], hence the memory state is terminated passively.
The results of $1000$ realizations of simulating Eq.~\eqref{eq:OU_process}, all with $\omega=1.006\cdot \omega_c$ but with different values of noise levels $\sigma$, are shown in Fig.~\ref{fig:noise_direct}~b. The average forgetfulness time $\av{T}$, of all realizations, shows an inverse relation to $\sigma$, seen by the dashed line.

Our LTP data analysis in Fig.~\ref{fig:LT}~d of young rats shows $\omega$ slightly above criticality (also $\omega=1.006\cdot \omega_c$), where the memory has been reported to vanish eventually after a week~\cite{lynch2004long}, much shorter than expected from LTM (lasting many years). Hence this case fits well to our definition. 
According to our findings of the $\omega$ values, we suggest that while aged rats could show deterministic plateau (slightly below the critical $\omega_c$, with $\omega=0.96\cdot \omega_c$, Fig.~\ref{fig:plateau}~d), young rats might show stochastic plateau ($\omega$ slightly above $\omega_c$, with $\omega=1.006\cdot \omega_c$ Fig.~\ref{fig:LT}~d). 

Theoretically, we can use the simulation results in Fig.~\ref{fig:noise_direct}~b to evaluate the neuronal noise $\sigma$. 
Assuming the termination of the LTP memory data after one week~\cite{lynch2004long}, i.e., $\av{T}\approx 10,000 [min]$, it yields according to Fig.~\ref{fig:noise_direct}~b a value of $\sigma/C \approx 0.05$. Thus, since for rats' current excitability threshold is $C=80 pA$~\cite{tateno2004threshold}, the average neuronal noise is $\sigma=0.05\cdot80=4 pA$. This value of $\sigma$ is in agreement with the estimated value of $~4.4pA$ current noise of a single neuron, as was suggested when investigating cultured neurons from rat hippocampus held near activation potential~\cite{diba2004intrinsic}.

In conclusion, we presented in the last two Sections, two possible scenarios for memory plateau formations and loss: (i) deterministic plateau: just below criticality (Fig.~\ref{fig:plateau}), the plateau emerges and drops fast when the current reaches from unstable state to the steady-state of $I=0$ (no-activity) in the Hopfield network model, 
or (ii) stochastic plateau: close and above criticality with noisy current fluctuations that at a certain time drops below criticality (Fig.~\ref{fig:noise_direct}~a), and terminates fast due to natural neuronal noise. 
Both options terminate 'red{persistent activity} passively and yield similar current trajectories, however, originate from two different mechanisms.

\section{Discussion}
In the present work we describe analytically the fundamental dynamics of memory persistent activity, by exploring the phase diagram under changes in the strength of the neuronal network synaptic connections and its effects on memory decay with time. 
Generally, our theory explains and shows, from self-sustained network state model perspective, that forgetfulness can be achieved in two ways, either passively via deterministic considerations that yield a plateau in the current as a result of being close but below criticality, or via stochastic noise perturbations that yield a plateau even at state just above criticality. Our analytical results suggest a new fundamental understanding of the memory termination. It ends spontaneously in contrast to the contemporary belief, which suggests that forgetfulness has an active mechanism (derives from inhibitor currents).

We note that at $\omega$ values far from criticality the active inhibitor currents might be relevant, but suggest here that at values near (above or below) criticality the memory-loss can be passive and spontaneously. 
Based on empirical LTP rats data, we also show that physiologically $\omega$ can indeed takes values near criticality.
Our results suggest that $\omega$ of young rats is just above criticality ($\omega>\omega_c$). Thus, it is quite interesting to find that, for the same experimental protocol, aged rats show values just below criticality ($\omega<\omega_c$), i.e., smaller than young rats. 
The relationship between LTP to memory was not fully understood, while memories last decades, LTP has been observed only for days (up to couple of weeks).	
Thus, excellent fit between LPT data and our simulations in Fig.~\ref{fig:plateau}~d, Fig.~\ref{fig:LT}~d and Fig.~\ref{fig:noise_direct}, supports our hypothesis that LTP could represent just a persistent activity which does not involve all of the LTM mechanisms changes.
We suggest that LTP does not evolve to long lasting memory since it produces too weak synaptic strength, with values only just above or just below criticality ($\omega_c$) for young and aged anesthetized rats respectively, which are probably not strong enough to produce substantial long lasting remodifications.

Note that the LTP is an artificial memory, which has some differences when compared to the naturally obtained (e.g., via visual stimulation) memory. 
Nevertheless, LTP is an excellent tool to study memory, in particularly using analytical models such as the Hopfield network model. 
Interestingly, our analysis and simulations results show that both our passive deterministic and passive stochastic plateau temporal values, as well as the typical time decaying exponents (see Eq.~S.23 and Eq.~S.28 (Supplementary)) depends on $\omega_c$ and $\tau$. This might explain the empirical findings of correlations~\cite{unsworth2010division, was2007reexamining} between all memory types.

Equation~\eqref{eq:Tplateau} and Fig.~\ref{fig:noise_direct} describe analytically and via simulations the typical time duration of the deterministic plateau and the stochastic plateau cases respectively.
While for the deterministic plateau case we offer deterministic stability approach, for the stochastic plateau case we offer a stochastic \textquoteleft first passage time\textquoteright{} mechanism. 
From Fig.~\ref{fig:noise_direct} we conclude that as the noise level $\sigma$ increases, the memory terminates earlier and there is a higher probability to return to the stable dormant state (no-memory state).
However, the opposite scenario, i.e., obtaining memory from the no memory state due to synaptic weight noise, is not possible according to our simulations. Adding synaptic noise to the model during the stable dormant state could not change the state, i.e., can not produce memory, even in proximity to the $\omega_c$ critical value. This points to a possible protection mechanism that prevents maintaining false memories, due to merely noise, at the long term (see also Supplementary Fig.~S5).

The neuronal noise is of clinical importance, that spans diverse clinical implications~\cite{dvir2018neuronal, dvir2019biased}. Our simulation results based on our framework for persistent activity in Fig.~\ref{fig:noise_direct}, suggest that medications for reducing neuronal noise level could prolong memory, and hence might therefore prevent fall risk in dementia patients\cite{sheridan2007role}. On the other spectrum of memory malfunctions, are diseases that are characterized by excessive memory, such as in post-traumatic stress disorder (PTSD). In those situations the patients might benefit from medications that promote forgetfulness by increasing synaptic noise. Generally, memory-loss is an important mechanism, allowing acquisition of new memories. Thus, our suggested passive forgetfulness mechanisms and analysis, could serve as an integral part of an healthy forgetfulness mechanism.

Several limitations of the present study, regarding the model simulations and the data interpretations should be noted. 
First, we compare our theoretical analyses and model simulations of persistent activity to experimental data as measured from EPSP of anesthetized rats~\cite{lynch2004long}. However, the relation between EPSP protocol of anesthetized rats and memory performances is not fully understood yet~\cite{lynch2004long}. 
Therefore, additional experimental platforms measuring memory should be compared to our model analysis.
\\
Second, using our model simulations, we adjust the model time constant parameter to $\tau=1$~[min], in accordance with sleep brain models~\cite{dvir2018neuronal}. This $\tau$ parameter main effect is on the time scaling of the current $I$ and EPSP, where smaller $\tau$ will achieve faster temporal dynamics.
For example, changing the cognitive state or the experimental stimulation protocol, e.g., wakening or natural stimulations, could affect $\tau$ and therefore change the time scales (faster or slower time dynamic).
Hence, future research should evaluate the $\tau$ values in different protocols.
\\
Finally, note that in our persistent activity model analysis, we assume that the neuronal network connections strength affects the membrane currents through a logarithmic function (Eq.~\eqref{eq:model}), as in earlier studies on memory~\cite{amit1995learning, durstewitz2000neurocomputational}. 
Although other functions can be used, for example hyperbolic tangent~\cite{lansner2009associative} or with a delay~\cite{leng2016basin}, we still expect similar conclusions without changing the main inferences, though the $\omega$-space diagram in Fig.~\ref{fig:phaseDiagram}, might be changed only quantitatively for other functions~\cite{sanhedrai2022reviving}.

\section*{Declaration of interests}
The authors declare no competing interests.

\section*{Funding}

We thank the Israel Science Foundation, the Binational Israel-China Science Foundation Grant No. 3132/19, the European Union under the Horizon Europe grant OMINO (grant number 101086321). 

H.S. acknowledges the support of the Presidential Fellowship of Bar-Ilan University,
Israel, and the Mordecai and Monique Katz Graduate Fellowship Program.

\section*{Data availability statement}

Data will be made available on request.

\clearpage
\newpage

%	\title{\huge  \color{blue}\textbf{-Supplementary-}\\
%	Mechanistic description of spontaneous loss of memory persistent activity based on neuronal synaptic strength}
	\begin{center}
		\color{blue} \bf \huge	
		\textbf{-Supplementary-}\\
		Mechanistic description of spontaneous loss of memory persistent activity based on neuronal synaptic strength
	\end{center}

\author{Hillel Sanhedrai$^{1}$, Shlomo Havlin$^{1}$ \& Hila Dvir$^{1*}$}
\date{}
\maketitle

\small{
	\begin{enumerate}
		\item
		\textit{Department of Physics, Bar-Ilan University, Ramat-Gan, Israel}			
	\end{enumerate}
	
	\begin{itemize}
		\item[\textbf{*}]
		\textbf{Correspondence}:\ \textit{dvirhila@gmail.com (HD)}
	\end{itemize} 
}

\vspace{4mm}

\setcounter{figure}{0}
\setcounter{equation}{0}
\setcounter{section}{0}
\renewcommand{\thefigure}{S\arabic{figure}}
\renewcommand{\theequation}{S.\arabic{equation}}
\renewcommand{\thesection}{S.\arabic{section}}

%\begin{center}
%	\color{blue} \bf \huge	
%	Supplementary
%\end{center}

%\vspace{-7mm}

\section{Intuitive explanation - analogy of block sliding down a slope}
\begin{figure} [h!]
	\centering
	\includegraphics[width=1.0\linewidth]{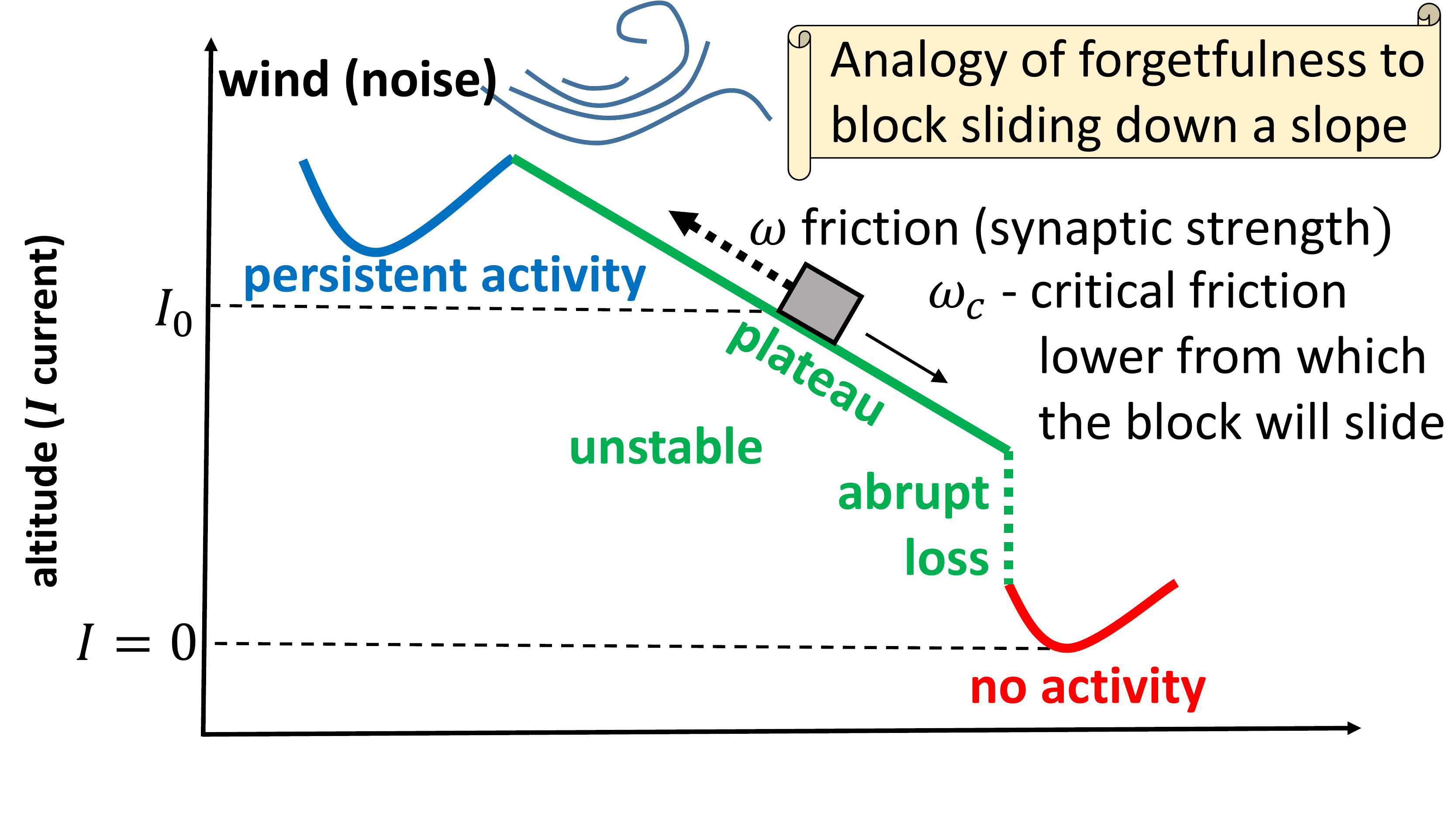}
	\caption{Illustration of the analogy to our suggested memory persistent activity forgetfulness mechanism.}
	\label{fig:intuitive_analogy}		
\end{figure}

%\vspace{-3mm}
We present here an intuitive explanation for our proposed forgetfulness mechanisms by using analogy to a simple model: block sliding down a slope. In this analogy the slope represents an unstable state and the end of the slope represents a stable state of a no-activity (of zero height, analogy to $I=0$), see Fig.~\ref{fig:intuitive_analogy}. The slope friction is analogous to $\omega$ and the block starting altitude is $I_0$. 
There exists a critical friction $\omega_c$ (criticality) below which the block will start to slide down towards zero altitude ($I=0$). 
Since the slope is an unstable plant, the block will eventually slide down towards the end of the slope (hence forgetfulness is unavoidable), however the block do spend some time on the slope, higher than the slope end altitude (having $I>0$, i.e., memory). Moreover, in case the friction $\omega$ is close to $\omega_c$ (small $\Delta\omega=\omega_c-\omega$), the block will reach the end (no-activity, i.e., no memory) after a longer time $t$, as demonstrated in Fig.~2~b-c. 
In case of a very high friction (close to $\omega_c$, criticality), the block will slide very slow, which is a close description to a plateau progression.

In section 3.1, 'Passive deterministic forgetfulness: Deterministic plateau', we show similar features  (though not exactly the same) that applies to the Hopfield network model, when we analyze the dynamics of Eq.~5 in case of transforming from unstable state near criticality (gray area in Fig.~2~a) to a stable forgetfulness.
According to the Hopfield model, Eq.~1 (or Eq.~5), two inverse forces control the dynamics of the ionic current: (i) The neural self-dynamics (the \textquoteleft $-I$\textquoteright~term), which is in accordance with the neuron basic Hodgkin-Huxley model~\cite{hille2001ion}, lead to current decay, and (ii) activation synaptic interactions (the \textquoteleft $\sum_{j \neq i} \m wij \ln \left( \frac{I_j}{C} \right)$\textquoteright~term). The self-dynamics in the block sliding illustration (Fig.~\ref{fig:intuitive_analogy}) is analogous to the gravitation force pushing the block downward, where the synaptic interactions is related to the opposed force of the friction between the block and the slope, resisting the sliding. Close to criticality the two forces are close to each other but the total membrane current $I$ will slowly decay, yielding a plateau shape of the membrane ionic current $I$.

In section 3.2.2, 'Passive stochastic forgetfulness: Stochastic plateau - effect of neuronal noise', we suggest an additional mechanism for forgetfulness based on neuronal noise. Continuing the analogy of block sliding down a slope, the neuronal noise is represented by wind. In case $\omega$ is close to $\omega_c$ and higher $\omega>\omega_c$, the block will not slide and stay at $I>0$, however, in conditions of winds (noise) there can be a scenario where the wind pushes the block, resisting the static friction. Hence, the block will slide to the end of the slope at $I=0$. Stronger winds (higher noise level $\sigma$) will yield shorter time duration at the slope, i.e., shorter $\av{T}$ in Fig.~4~b.

\section{Mean-field approximation}

Starting with Eq.~1 in the main text,
\begin{equation}
\tau \dod{I_i}{t} = -I_i + \sum_{j \neq i} \m wij \ln \left( \frac{I_j}{C} \right) \theta\left(\frac{I_j}{C}-1 \right) + I_{\rm Aff},
\label{eq:modelSI}
\end{equation}
we use a mean-field approach~\cite{gao2016universal} and assume that since the summation is on large number of variables from a narrow distribution due to their similar degrees and weight distribution, we can replace the variables $I_j$ by their average over the entire network, $I$, yielding
\begin{equation}
\tau \dod{I_i}{t} = -I_i + \sum_{j \neq i} \m wij \ln \left( \frac{I}{C} \right) \theta\left(\frac{I}{C}-1 \right) + I_{\rm Aff},
\label{eq:MF1SI}
\end{equation}
where the average current is simply
\begin{equation}
I = \frac{1}{N}\sum_{i =1}^{N} I_i.
\label{eq:ISI}
\end{equation}
Next, we operate an average on both sides of Eq.\ \eqref{eq:MF1SI},
\begin{eqnarray}
\frac{1}{N}\sum_{i =1}^{N} \tau \dod{I_i}{t} =\frac{1}{N}\sum_{i =1}^{N} (-I_i) + \frac{1}{N}\sum_{i =1}^{N}\sum_{j \neq i} \m wij \ln \left( \frac{I}{C} \right) \theta\left(\frac{I}{C}-1 \right) + I_{\rm Aff},
\label{eq:MF2SI}
\end{eqnarray}
by which we obtain Eq.~2 in the main text,
\begin{equation}
\tau \dod{I}{t} = -I + \omega (N-1) \ln \left( \frac{I}{C} \right)\theta\left(\frac{I}{C}-1 \right) + I_{\rm Aff},
\label{eq:MFSI}
\end{equation}
where 
\begin{equation}
\omega = \frac{1}{N(N-1)}\sum_{i =1}^{N} \sum_{j \ne i} w_{ij}
\label{eq:wSI}
\end{equation}
is the average weight.

\begin{figure}[h]
	\centering
	\includegraphics[width=0.99\linewidth]{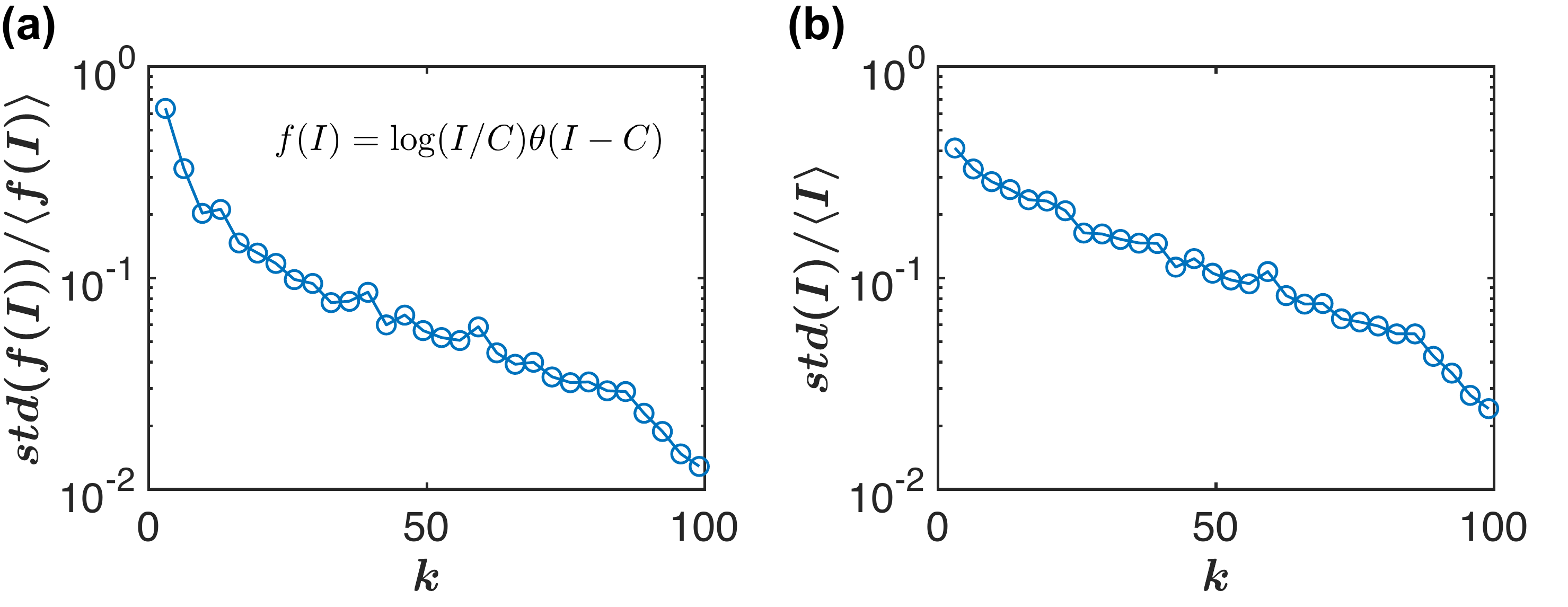}	
	\caption{\footnotesize{\bf Mean-field approximations validation.}
		In our theory, we replaced a varying quantity by a constant average quantity by two steps as follows, $f(I_j)=\log(I_j/C)\theta\left(\frac{I_j}{C}-1 \right)\approx \av{f(I)}\approx f(\av{I})$. To validate each one of the steps we measured the standard deviation of these variables. We measured the steady state of the system for increasing average degree, $k$, and for this state we measured the standard deviation relative to the average. For different values of $k$, we need a corresponding value of $\omega$ to get the non-zero steady state, thus we set $\omega k=7$. (a) The relative standard deviation of $f(I_j)$ decays to small values for a densely connected graph which is discussed in this manuscript. This justifies the approximation $f(I_j)\approx \av{f(I)}$. (b) The relative standard deviation of $I_j$ decays to small values for a densely connected graph which is discussed in this manuscript. This justifies the approximation $\av{f(I)}\approx f(\av{I})$. 
	}
	\label{fig:MFvalidation}
\end{figure}

In Fig.\ \ref{fig:MFvalidation} we explore and validate the mean-field approximation done in Eq.\ \eqref{eq:MF1SI} resulted in Eq.\ \eqref{eq:MFSI}. We measure, in simulation, the term of interaction for each node, $ f(I_j) = \ln \left( I_j/C \right) \theta\left(\frac{I_j}{C}-1 \right)$, and the standard deviation of this quantity, to show that when the degree is large then the fluctuations become negligible, and therefore the approximation of replacing this quantity by its average is justified and provides an accurate prediction. In addition, to examine the second mean-field approximation we use, that of inserting the averaging into the functions, $\av{f(I)}=f(\av{I})$, we measure the standard deviation of $I$, and show that it becomes smaller for higher degree what makes the approximation acceptable.  
To this end, we set a various values for the average degree $k$ (number of connections of a node), ranged from $k=3$, representing a sparse network, to $k=N-1$ which is a complete graph that was considered in all simulations in this manuscript. The state of the system was measured after approaching the steady state located above criticality in the persistent activity regime. As shown in Figs.\ \ref{fig:MFvalidation}a,b the standard deviation becomes small for high degree, what explains the good agreement of our theory with the simulation results along the whole manuscript. Note that the approximation is even better since the term $ f(I_j) = \ln \left( I_j/C \right) \theta\left(\frac{I_j}{C}-1 \right)$ does not appear as a single term but rather within a sum over all neighbors, i.e.\ as an average over the neighbors, and this average has a smaller relative standard deviation by factor of $1/\sqrt{k}$, following the rule of averaging of random variables.

\vspace{25mm}

\section{Fixed points and stability analysis}

To find the fixed points of the system including both stable and unstable states, we just demand relaxation in Eq.~5 in the main text, i.e. $\dif I/\dif t=0$,
\begin{equation}
0 = -I + \omega (N-1) \ln \left( \frac{I}{C} \right)\theta\left(\frac{I}{C}-1 \right) ,
\label{eq:fpSI}
\end{equation}
whose solutions are the fixed points.

\begin{figure}
	\centering
	\includegraphics[width=0.5\linewidth]{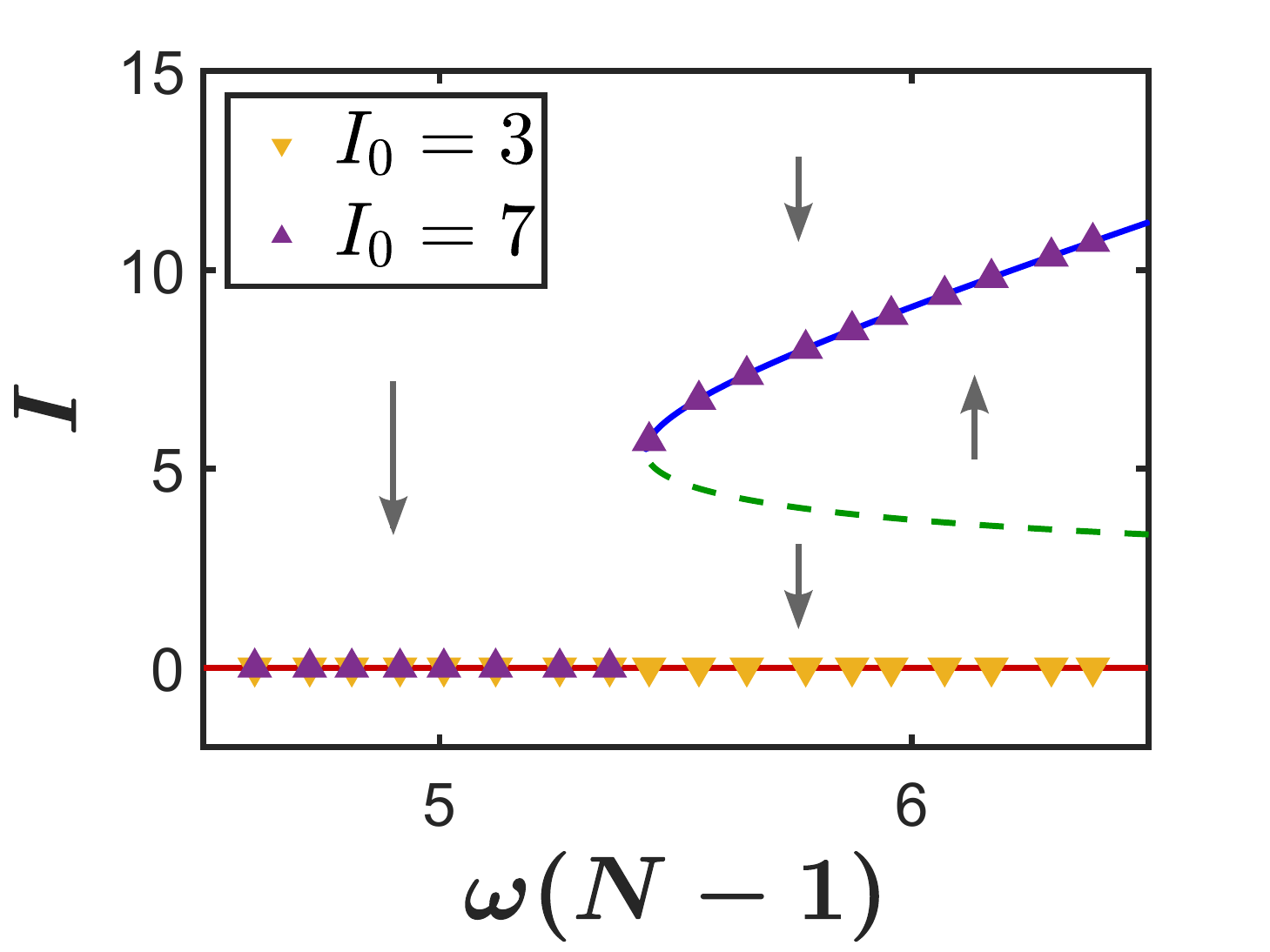}	
	
	\caption{\footnotesize{\bf Memory phase diagram.}
		The phase diagram in $\omega$-space yielded by Eqs.~\eqref{eq:MFSI},\eqref{eq:fpSI},\eqref{eq:StabCondSI}. The blue and the red continuous lines represent the steady states, while the dashed green line represents an unstable state separating between the two basins of attraction.  
		Simulation (symbols) with normally distributed synaptic weights with $\sigma=\omega_c/4$ starting with high (purple) and low (orange) initial conditions are shown for $N=100$, $\tau=1$ and $C=2$.
	}
	\label{fig:phaseDiagramSI}
\end{figure}

To classify whether a solution of Eq.\ \eqref{eq:fpSI} is stable, we demand the known stability condition \cite{strogatz2018nonlinear},
\begin{equation}
\pd{}{I} \left( \dod{I}{t} \right) < 0.
\label{eq:StabCondSI}
\end{equation}
One can see that the fixed point $I=0$ is stable since the term in Eq.\ \eqref{eq:StabCondSI} is $-1<0$. There are also positive stable fixed points,  see Fig.\ \ref{fig:phaseDiagramSI}. In between, there exist unstable fixed points (dashed line) separating between the basins of attraction of the stable states.

\section{Tipping point}

To find the critical (tipping) point ($\omega_c,I_c$) of the transition from no-activity to persistent activity, shown in Fig.\ \ref{fig:phaseDiagramSI}, we notice that this point is a junction of a stable fixed point (continuous blue line in Fig.\ \ref{fig:phaseDiagramSI}) and unstable one (dashed green line), resulting in vanishing of the term in Eq.\ \eqref{eq:StabCondSI}. 
Therefore, two conditions should be satisfied in the critical point: 1.\ fixed point, 2. a transition from unstable to stable, as follows,
\begin{equation}
\begin{array}{ll}
1. &  \dod{I}{t} = 0,
\\[10 pt]
2. & \dfrac{\partial}{\partial I} \left( \dod{I}{t} \right) = 0.
\end{array}
\label{eq:twoConditions}
\end{equation}
Using Eq.~5 in the main text, we obtain,
\begin{equation}
\begin{array}{ll}
1. & -I_c + \omega_c (N-1) \ln \left( \dfrac{I_c}{C} \right) = 0,
\\[10 pt]
2. & -1 + \omega_c (N-1)\dfrac{1}{I_c} = 0.
\end{array}
\label{eq:twoConditions2}
\end{equation}
The solution is
\begin{equation}
\omega_c = \dfrac{e\cdot C}{N-1},
\label{eq:wc}
\end{equation}
and 
\begin{equation}
I_c=\omega_c (N-1)=e\cdot C,
\label{eq:Ic}
\end{equation}
as given in Eq.~6 in the main text.

\section{Persistent activity typical relaxation time}

To get the current exponential decay for $\omega>\omega_c$, in the limit of $I \to I_{\rm LT}$, we expand Eq.~5 in the main text as $I=I_{\rm LT}+\delta I$ where $\delta I$ is small, and obtain,
\begin{equation}
\tau \dod{\delta I}{t} = -I_{\rm LT}-\delta I + \omega (N-1) \ln \left( \frac{I_{\rm LT}+\delta I}{C} \right) ,
\end{equation}
which can be written as
\begin{equation}
\tau \dod{\delta I}{t} = -I_{\rm LT}-\delta I + \omega (N-1) \bigg[ \ln \left( \frac{I_{\rm LT}}{C} \right) +\ln \left(1+ \frac{\delta I}{I_{\rm LT}} \right)  \bigg] .
\end{equation}
Using Eq.~11 in the main text, and expanding $\ln(1+\epsilon)$ for small $\delta I$ (Taylor series), we obtain
\begin{equation}
\tau \dod{\delta I}{t} \approx - \left( 1 - \frac{\omega (N-1)}{I_{\rm LT}} \right) \delta I - \frac{1}{2}\omega (N-1) \left( \frac{\delta I}{I_{\rm LT}} \right)^2 .
\end{equation}
Neglecting the second order term (which is equivalent to assuming $\delta I / I_{\rm LT} \ll \ln (I_{\rm LT}/I_c)$), and additionally using Eq.~11 again, we obtain
\begin{equation}
\tau \dod{\delta I}{t} \approx -\bigg( 1 -\frac{1}{\ln (I_{\rm LT}/C)} \bigg) \delta I .
\end{equation}
This implies an exponential decay with typical time decay
\begin{equation}
\tau_{\rm LT} = \frac{\tau}{1 -\frac{1}{\ln (I_{\rm LT}/C)}} ,
\label{eq:tauLTSI}
\end{equation}
which is Eq.~13 in the main text.

\subsection{Close to criticality}

Next, we evaluate $\tau_{\rm LT}$ when $\omega$ is close to the critical value $\omega_c$ from above. To track the behavior in proximity to $\omega_c$ (for small $\Delta\omega$), we expand Eq.~11 by $\omega=\omega_c+ \Delta\omega$ and $I_{\rm LT} = I_c + \delta I_{\rm LT}$. Recalling $I_c=e\cdot C, \ \omega_c=e\cdot C/(N-1)$ (Eq.~6), we get
\begin{equation}
0 = - e\cdot C - \delta I_{\rm LT} + (e\cdot C + \Delta\omega (N-1)) \ln (e+\delta I_{\rm LT}/C) ,
\end{equation}
which is
\begin{equation}
0 = -I_c - \delta I_{\rm LT} + (I_c + \Delta\omega (N-1)) \left[1+\ln \left(1+\frac{\delta I_{\rm LT}}{I_c}\right)\right].
\end{equation}
Expansion of $\ln(1+\epsilon)$ for small $\epsilon$ gives
\begin{equation}
0 \approx -I_c - \delta I_{\rm LT} + (I_c + \Delta \omega (N-1)) \left[1+\frac{\delta I_{\rm LT}}{I_c} - \frac{\delta I_{\rm LT}^2}{2I_c^2} + \dots \right].
\end{equation}
This yields
\begin{equation}
\delta I_{\rm LT} \approx I_c \sqrt{2 \Delta\omega/\omega_c} .
\end{equation}
Substituting this into Eq.~13, we finally get
\begin{equation}
\tau_{\rm LT} = \frac{\tau}{1 -\frac{1}{\ln (e+\delta I_{\rm LT}/C)}} \approx \frac{\tau}{1 -\frac{1}{1+\delta I_{\rm LT}/(e\cdot C)}} \approx \frac{\tau}{\delta I_{\rm LT}/I_c} .
\end{equation}
Hence, close to criticality, $\omega\to\omega_c$ (though still $\omega>\omega_c$),
\begin{equation}
\tau_{\rm LT}  \approx \frac{\tau}{\sqrt{2 \Delta\omega/\omega_c}}, 
\label{eq:tauLTcloseIS}
\end{equation}
which is Eq.~14 in the main text.

\section{No-activity typical time decay}

\begin{figure}
	\centering
	\includegraphics[width=0.9\linewidth]{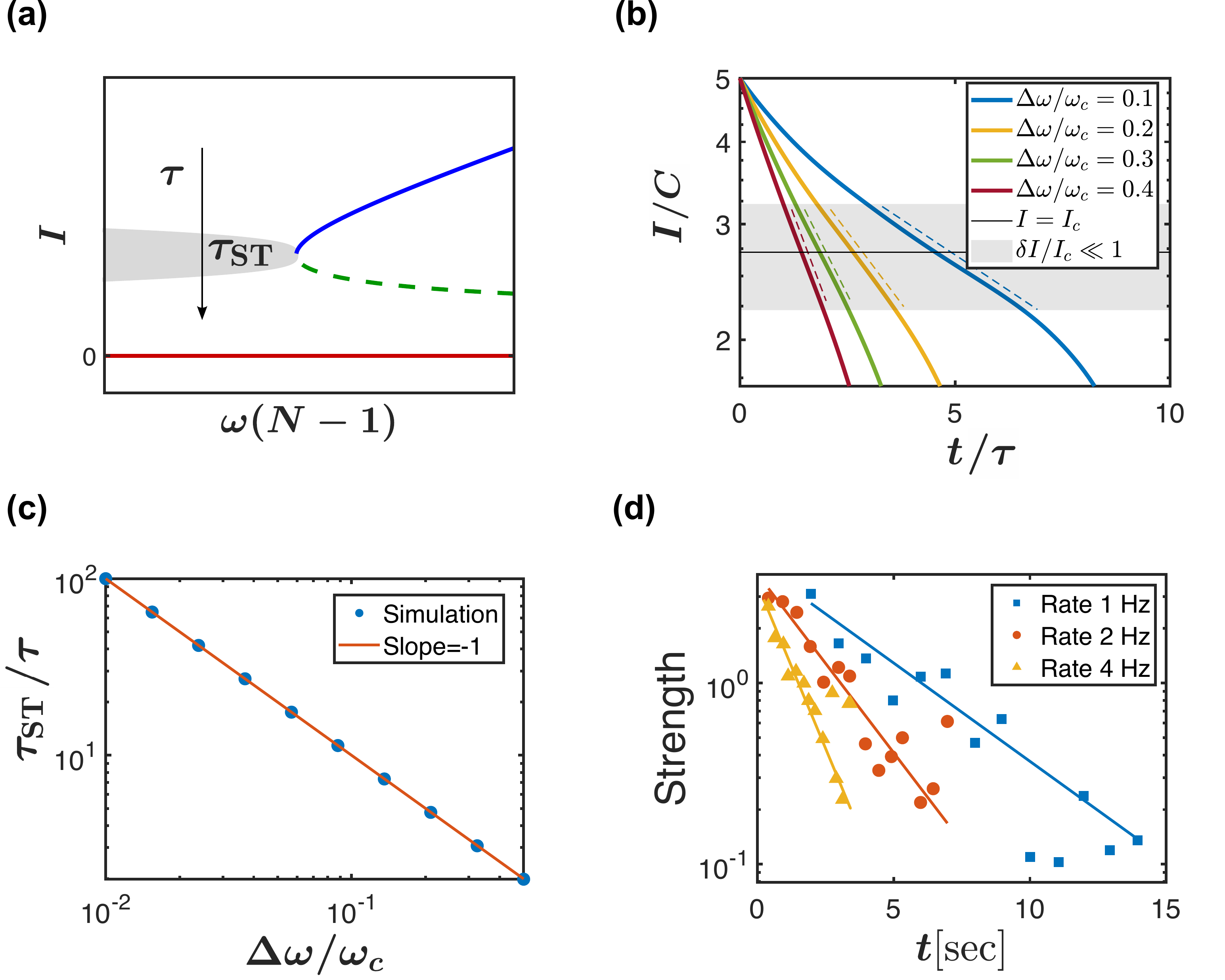}		
	
	\caption{\footnotesize{\bf Short-term memory temporal decay.}  
		(a) The current decays to zero because the weight is below criticality, $\omega<\omega_c$, thus the system is in the dormant region. The regime where $\tau_{\rm ST}$ is relevant is marked in gray. 
		(b) Simulation results with four different $\Delta\omega/\omega_c$ values marked in colored lines. The gray area around $I_c$ marks the range where the decay of $I$ in time is exponential with typical time $\tau_{\rm ST}=\tau/(\Delta\omega/\omega_c)$ (dashed lines), which is in good agreement with theory, Eq.\ \eqref{eq:tauSTSI}. Note that we show $I$ only up to the threshold value $C$ since below it there exists no relevant signal. 
		(c) It is seen that $\tau_{\rm ST}$ from the simulated curves (like in (b)) scales as $\Delta \omega ^{-1}$, which supports the theory, Eq.\ \eqref{eq:tauSTSI}. The simulations results in panels (b)-(c) were obtained based on a network of $N=100$ neurons, $C=2$ and $\tau=1$~min.
		(d) Recalling of a limited list is a common procedure to quantify short-term-memory (STM) empirically. The ability to recall words/numbers from a list depends also on the test itself. Here we show an example from empirical tests~\cite{wickelgren1970time} for several word presenting rates. Comparing this panel to panel (b) shows how STM dynamics across different rates can be modeled by using different synaptic connection strengths.
		The lines are the analytical theory fittings of the empirical data with time decays $\tau=4,2.2,1.2$~min.
		Note that towards 10sec, the blue dots (rate of 1Hz) markedly reach below the exponential fitting blue line, which has some resemblance to the curved lines in panel (b) when $I$ reaches below the valid gray area (mostly noted in the blue symbols in panel (b)).
	}
	\label{fig:ST}
\end{figure}

For exploring the exponential decay in $\omega<\omega_c$ region, we analyze the evolution of $I(t)$ around $I_c$. Therefore, we define here $\omega=\omega_c-\Delta\omega$ (this is in contrast to the persistent activity where $\omega>\omega_c$ which defines $\omega=\omega_c+\Delta\omega$).
Let us expand Eq.~5 in the main text at $I=I_c+\delta I$, for small deviation $\delta I$,
\begin{equation}
\tau \dod{I}{t} = -I + (\omega_c-\Delta \omega) (N-1) \ln \left( \frac{I_c+\delta I}{C} \right),
\end{equation}
that is,
\begin{equation}
\tau \dod{I}{t} = -I + (\omega_c-\Delta \omega) (N-1) \left[\ln \left( \frac{I_c}{C} \right) + \ln \left( 1 + \frac{\delta I}{I_c} \right) \right] .
\end{equation}
Recalling $I_c=e\cdot C$ and $\omega_c=e\cdot C/(N-1)$, Eq.~6, and expanding the Taylor series of $\ln(1+\epsilon)$ to the first order, we get
\begin{equation}
\tau \dod{I}{t} \approx -I + (I_c-\Delta \omega (N-1) ) \left[1 + \frac{\delta I}{I_c} -\frac{1}{2}\left(\frac{\delta I}{I_c}\right)^2\right] \approx - \frac{\Delta\omega}{\omega_c} I ,
\label{eq:dIdtAppCloseIcSI}
\end{equation}
where we neglect the second order term, \textit{i.e.}\ we assume $ (\delta I / I_c)^2 \omega (N-1) \ll (I_c+\delta I)\Delta\omega/\omega_c $. 
Thus, this region gets smaller close to criticality $\omega_c$, as illustrated in Fig.\ \ref{fig:ST}~a in the main text (this region is marked in gray). 

Finally, Eq.\ \eqref{eq:dIdtAppCloseIcSI} yields 
\begin{equation}
I(t) \sim \exp (-t/\tau_{\rm ST}),
\end{equation}
where the time decay is
\begin{equation}
\tau_{\rm ST} = \frac{\tau }{\Delta\omega/\omega_c} ,
\label{eq:tauSTSI}
\end{equation}

\section{Memory creation}

\begin{figure}
	\centering
	\includegraphics[width=0.9\linewidth]{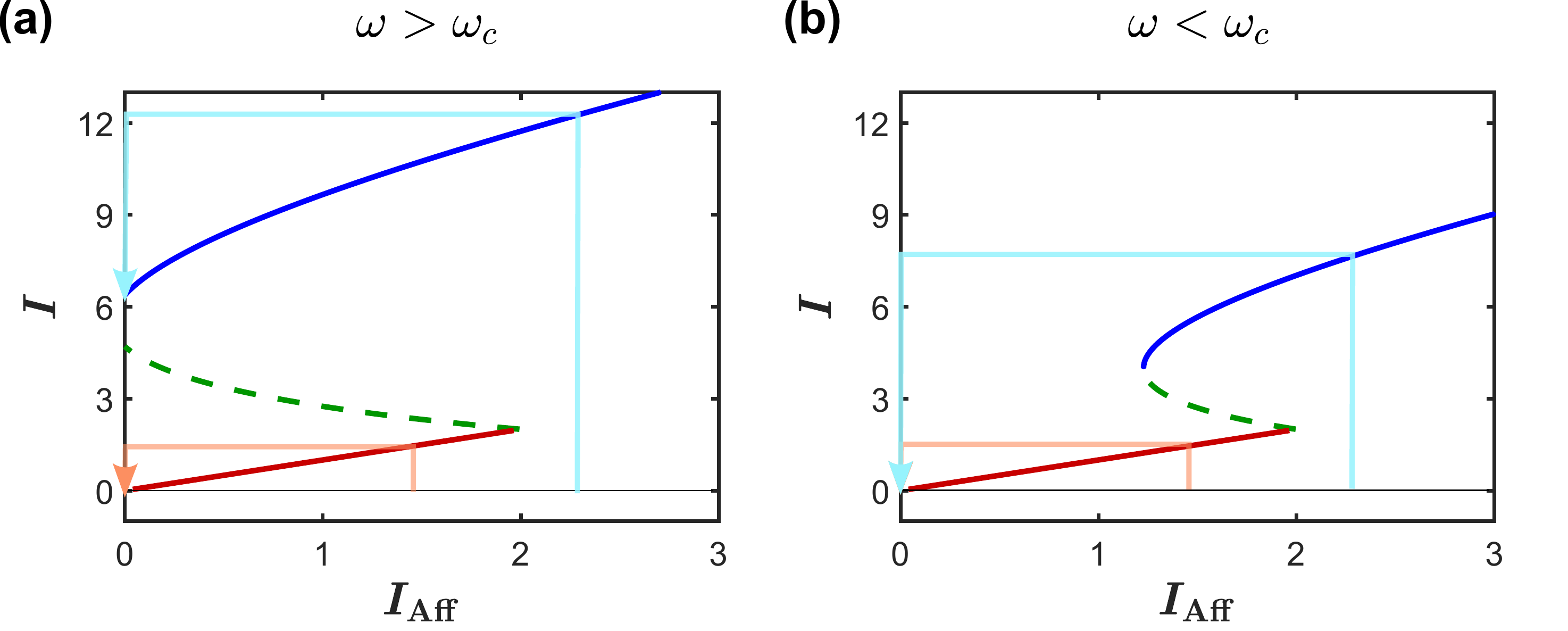}	
	\caption{\footnotesize\footnotesize{\bf No creation of memory via noise.} 
		While noise in the weights greatly affect memory loss (change the state from memory state $\omega>\omega_c$ to no memory state - see Fig.~4), the appearance of noise in the no memory state ($\omega<\omega_c$) cannot change the state toward memory. This effect can be understood from the illustration of the states with (a) $\omega>\omega_c$ and (b) $\omega<\omega_c$. Whereas changing $\omega$ greatly affects the stable memory state (in blue) and the left transition point, the stable no-memory state (in red) and the right transition point remain unaffected. Hence, variations in $\omega$ cannot affect the no-memory stable state.
	}
	\label{fig:creationSI}
\end{figure}

Figure~\ref{fig:creationSI} demonstrates, based on Eq.~2 in the main text, the possibility of creation memory. The memory is generated due to the contemporary existence of $I_{\rm Aff}>C$ which causes to the increasing of $I$. Once the stimulation finished and $I_{\rm Aff}$ vanished, the current decays either to its long-term memory steady state or to zero in the case of short-term memory as shown in Fig.~1. One can see in Fig.\ \ref{fig:creationSI}a that for large weight, $\omega>\omega_c$, and $I_{\rm Aff}>C$, \textit{i.e.}\ right to the point at which the curve gets broken, a long-term memory is created (light blue arrow). This is because the current increases much during the stimulation, and once $I_{\rm Aff}$ becomes zero, the current decreases to the high stable state (blue), and stays there.
In contrast, if $I_{\rm Aff}<C$ (orange arrow), the current increases slightly to the low stable state (red), such as when $I_{\rm Aff}=0$, the current decays to zero, and no memory has been created.
While for $\omega<\omega_c$, as shown in Fig.\ \ref{fig:creationSI}b, even if the stimulation is strong (light blue arrow), no long-term memory is created since, there is no high stable state for $I_{\rm Aff}=0$, thus the current decays to zero. 

However, note that the right tipping point (intersection of red and green lines) does not change with $\omega$. Thus, the condition of a major stimulation is always $I_{\rm Aff}>C$ independent on $\omega$. The weight $\omega$ impacts only the ability of the system to preserve the memory for a long term. Therefore, a noise in the weights can not cause an emergence of a memory although it can cause a loss of a memory as shown in Fig.~4.

\end{document}